\documentclass{article}
\usepackage{spconf,graphicx}
\usepackage[utf8]{inputenc}
\usepackage[nolist]{acronym}

\usepackage{bm}
\usepackage{mathtools}
\usepackage{physics, amsmath,amssymb,amsfonts}
\usepackage{IEEEtrantools}
\usepackage{microtype}
\usepackage[dvipsnames, table]{xcolor}
\usepackage{booktabs,dcolumn, multirow}
\usepackage{dirtytalk}
\usepackage{cite}
\usepackage{placeins}
\usepackage[subtle]{savetrees}
\usepackage[hidelinks]{hyperref}
\usepackage{diagbox}

\usepackage[caption=false]{subfig}
\usepackage{tikz}
\usetikzlibrary{quotes,arrows.meta, calc, patterns.meta, shapes.multipart, shapes.arrows, shapes.misc, spy, decorations.pathreplacing, arrows.meta, backgrounds, shapes.geometric, intersections, matrix}
\usepackage{pgfplots}
\pgfplotsset{compat=1.15}
\usepgfplotslibrary{colormaps,groupplots,dateplot}
\usepackage{booktabs}

\let\OLDthebibliography\thebibliography
\renewcommand\thebibliography[1]{
  \OLDthebibliography{#1}
  \setlength{\parskip}{0pt}
  \setlength{\itemsep}{0.5pt plus 2ex}
}

\title{Spatially Selective Deep Non-linear Filters for Speaker Extraction}
\name{Kristina Tesch and Timo Gerkmann}
\address{
  Signal Processing (SP), Universität Hamburg, Germany\\kristina.tesch@uni-hamburg.de, timo.gerkmann@uni-hamburg.de}

\begin{acronym}
\acro{DFT}{discrete Fourier transform}
\acro{MVDR}{minimum variance distortionless response}
\acro{PDF}{probability density function}
\acro{MMSE}{minimum mean square error}
\acro{ML}{maximum likelihood}
\acro{SNR}{signal-to-noise ratio}
\acro{MAP}{maximum a posteriori}
\acro{ASR}{automatic speech recognition}
\acro{POLQA}{perceptual objective listening quality analysis}
\acro{MOS}{mean opinion score}
\acro{PESQ}{perceptual evaluation of speech quality}
\acro{EM}{expectation maximization}
\acro{DNN}{deep neural network}
\acro{LSTM}{long short-term memory}
\acro{FF}{feed-forward}
\acro{cIRM}{complex ideal ratio mask}
\acro{IRM}{ideal ratio mask}
\acro{STFT}{short-term Fourier transform}
\end{acronym}

\def\vec#1{\ensuremath{\mathbf{#1}}}

\newcommand{\vecY}{\vec{Y}}

\newcommand{\vecV}{\vec{V}}

\newcommand{\vecX}{\vec{X}}

\begin{document}
\ninept
\maketitle
\begin{abstract}
In a scenario with multiple persons talking simultaneously, the spatial characteristics of the signals are the most distinct feature for extracting the target signal. In this work, we develop a deep joint spatial-spectral non-linear filter that can be steered to an arbitrary target direction. For this we propose a simple and effective conditioning mechanism, which sets the initial state of the filter's recurrent layers based on the target direction. We show that this scheme is more effective than the baseline approach and increases the flexibility of the filter at no performance cost. The resulting spatially selective non-linear filters can also be used for speech separation of an arbitrary number of speakers and enable very accurate multi-speaker localization as we demonstrate in this paper. 
\end{abstract}
\section{Introduction}
\label{sec:intro}
{\let\thefootnote\relax\footnote{\footnotesize This work was funded by the Deutsche Forschungsgemeinschaft (DFG, German Research Foundation) — project number 508337379. We thank Rohde\&Schwarz SwissQual AG for their support with POLQA.}}
In our everyday life, we are often confronted with the task of listening to a target speaker in a challenging acoustic environment containing noise, interfering human speakers, and reverberation. It is widely known that humans are able to utilize spatial information perceived with both ears to draw attention towards a particular direction of interest. Similarly, spatial information can be used in addition to tempo-spectral information for target speaker extraction in many applications since devices like hearing aids, video-conferencing systems or voice-controlled assistants are nowadays commonly equipped with multiple microphones. 

Research into spatial filtering has a long-standing history, which has led to the traditional beamformers, e.g., the delay-and-sum \cite{vary2006digital} or \ac{MVDR} beamformer \cite{vary2006digital, doclo2015assistedlistening}. While \acp{DNN} are considered the state-of-the-art in single-channel speech enhancement and separation, their integration into multi-channel techniques is a very active field of research. Here, one of the most influential ideas of the last years was to use neural networks for beamformer parameter estimation \cite{heymann2015blstm, xiao2016deepbeamforming}. Despite ease of use and demonstrated robustness of this method, the main drawback of using \acp{DNN} only for parameter estimation is that the limitations of the linear beamforming model cannot be overcome, nor can we benefit from joint processing of spatial and tempo-spectral information.

In contrast, an increasing number of recent works, trains a \ac{DNN}-based filter to perform multi-channel speech enhancement, speaker extraction or separation directly with promising results \cite{tolooshams2020cadunet, 2020liNarrowbandDeepFiltering, tan2022spatiospectralfilter, li2022eabnet, halimeh2022cospa, markovic2022nsfmeta}. The theoretic foundation for the potential performance improvements of \ac{DNN}-based multichannel filters over traditional or DNN-driven beamforming and postfiltering is layed out in our prior work \cite{tesch2021nonlinearspatialfilteringtasl}. By means  of statistical derivations and proof-of-concept experiments we have shown that (1) a linear beamformer will deliver optimal performance only in rare cases, namely under a multi-variate Gaussian noise assumption, and (2) that non-linear joint spatial-spectral filters may drastically outperform the beamforming plus postfiltering schemes in other cases. DNNs are a natural choice to implement such non-linear joint spatial-spectral filters for practical applications. %

Consequently, we \cite{tesch2022interspeech, tesch2022tasl}, and also others \cite{markovic2022nsfmeta}, have shown that such a DNN-based joint spatial and tempo-spectral non-linear filter drastically outperforms an oracle \ac{MVDR} beamformer followed by a single-channel post-filter. For this, we evaluated on a speaker extraction task with five interfering speakers. Part of the speaker extraction task is to identify the target speaker. In the literature, different cues have been investigates for this, e.g., enrollment utterances \cite{delcroix2018speakerbeam, wang2019voicefilter} and video information \cite{afouras2018conversation, michelsanti2021audiovisual}. 

In this work, however, we focus on the spatial location of the target speaker as cue. Many previous works have explored using spatial features to aid speech separation or speaker extraction \cite{gu2019neural, gu2021complexnsf, chen2018anglefeature, wang2018mcdeepclustering, wang2019combining}. For example, Gu et al. \cite{gu2019neural, gu2021complexnsf} and Chen et al. \cite{chen2018anglefeature} have introduced so-called directional features into their speech separation and extraction systems, which indicate time-frequency bins that are dominated by signal components arriving from a particular direction and are used as additional inputs besides tempo-spectral features. Markovi\'c et al. \cite{markovic2022nsfmeta} follow a different approach and define spatial regions, e.g., left and right, and train a non-linear filter that suppresses signals from the undesired region but not from the desired region. Tan et al. \cite{tan2022spatiospectralfilter} train a non-linear spatial filter that implicitly steers towards the speech source in enhancement tasks and learns to resolve the speaker-permutation problem by implicitly sorting the speaker outputs according to their location. 

In contrast, in this work, we aim for a non-linear joint filter that can be flexibly steered in a direction of choice. This is a major improvement in comparison with our previous well-performing filter \cite{tesch2022interspeech, tesch2022tasl}, which is restricted to a fixed look-direction and thus requires re-training for other directions. For this, we propose a simple conditioning mechanism based on an angular grid with $2^\circ$ resolution. In comparison with the implicit conditioning mechanism proposed in \cite{jenrungrot2020coneofsilence}, which manipulates the input signal, our proposed conditioning scheme is more explicit and does not make a far-field assumption. 

The rest of this paper is structured as follows: We formally define the speech extraction problem in Section \ref{sec:probdef} and explain the non-linear filter and its conditioning on a target direction in Section \ref{sec:jnf}. Section \ref{sec:expsetup} describes the experimental setup including datasets and in Section \ref{sec:results}, we present results on the effectiveness of the conditioning mechanism and the spatial selectivity of the resulting filter.

\section{Problem definition}\label{sec:probdef}
This work targets the so-called cocktail-party problem: extracting the speech signal uttered by a target speaker from interfering speech. We assume that the corrupted signal is captured by a microphone array with $C$ channels and denote with $x_\ell(t)$ the recording of the target speech signal $s(t)$ obtained by the $\ell$'s microphone. The time-domain signal $x_\ell(t)$ is not only a time-shifted version of $s(t)$ caused by the propagation delay between the speaker and the microphone but also includes reverberation resulting from reflections of the signal from the surrounding walls. 

We apply the \ac{STFT} to obtain a frequency-domain representation $X_\ell(k,i)\in\mathbb{C}$ with frequency-bin index $k$ and time-frame index $i$. The spectral coefficients for all channels are stacked into a vector $\vecX(k,i) = [X_0(k,i), ...,X_{C-1}(k,i)] \in\mathbb{C}^C$. We employ the same signal model to model interfering speech signals and denote the STFT representation of the sum of all interfering signals as $\vecV(k,i)$. By the additive signal model, the noisy target signal, $\vecY(k,i)$, corrupted by interfering speakers, is then given by the sum of the target signal and interfering signal, i.e., 
\begin{equation}
    \vecY(k,i) = \vecX(k,i) + \vecV(k,i).
\end{equation}

Given the noisy recording $\vecY(k,i)$ we aim to recover the clean target speech signal $S(k,i)$ except for a time-shift caused by the propagation delay to the chosen reference microphone, for which we pick the first channel. 

\section{Spatially selective non-linear filter}\label{sec:jnf}

In our previous work \cite{tesch2022interspeech, tesch2022tasl}, we have shown that a DNN-based non-linear filter that jointly performs spatial and tempo-spectral filtering, can implicitly be steered into a specific direction, when trained on a fixed geometric setting. Here, we extend the joint non-linear filter from \cite{tesch2022interspeech, tesch2022tasl}, displayed on the left side of Figure \ref{fig:netarc}, with a conditioning mechanism, shown on the right side of Figure \ref{fig:netarc}, that allows the filter to be flexibly steered in a desired direction. 

\subsection{Joint spatial and tempo-spectral non-linear filter}

As indicated by the top left yellow box, the filter takes the frequency-domain raw multi-channel observations as input. Including the batch dimension denoted by $B$, the input is four-dimensional with $T$ being the number of time-steps, and $F$ the number of the frequency-bins. The real and imaginary parts for all $C$ microphone channels are stacked resulting in the last dimension being $2C$. The filter is composed of only three layers represented by dark green boxes and outputs an estimate of a compressed \ac{cIRM}. We use compression parameters $\mathcal{K}=\mathcal{C}=1$ as defined in \cite{2016williamsonComplexRatioMasking} that comply with the range of the $\tanh$ activation function used in the last layer. The estimate of the target speech signal $\hat{S}(k,i)$ is then obtained by multiplying the uncompressed mask $\mathcal{M}(k,i)\in\mathbb{C}$ with the reference channel's noisy recording $Y_0(k,i)$, i.e., 
\begin{equation}
    \hat{S}(k,i) = \mathcal{M}(k,i)\cdot Y_0(k,i).
\end{equation}

\begin{figure}
    \centering
    \includegraphics{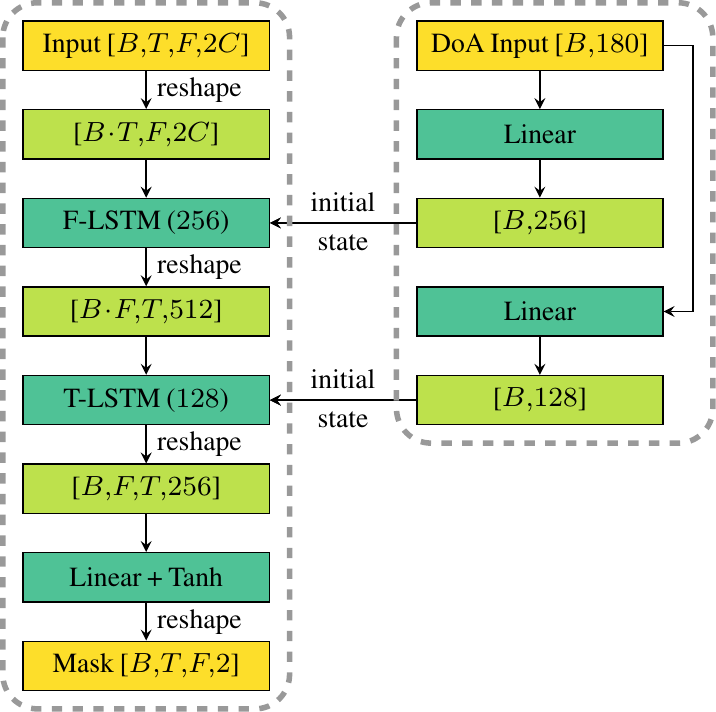}
    \vspace{-0.5em}
    \caption{Illustration of the network architecture. The left part shows the mask estimation network that performs joint spatial and tempo-spectral filtering and the right part shows the conditioning mechanism that enables the filter to be steered towards a chosen direction.}
    \vspace{-0.5em}
    \label{fig:netarc}
\end{figure}

The network design is inspired by the work of Li and Horaud \cite{li2019narrowband}, who proposed a narrow-band multi-channel speech enhancement scheme. Their core idea is to use a simple network structure (two bi-directional \ac{LSTM} layers and one linear layer) and process all frequency-bins independently while sharing the network parameters between all frequencies. This processing scheme puts a focus on spatial and temporal information and neglects the information present in the frequency dimension. However, our previous work \cite{tesch2022interspeech, tesch2022tasl} has shown that spectral information, including the correlations between neighboring frequency-bins, should be included in the processing to obtain a filter with high spatial selectivity. Therefore, we rearrange the data such that the first \ac{LSTM} layer (F-LSTM) focuses on spatial and spectral information and the second \ac{LSTM} layer (T-LSTM) focuses on spatial and temporal information. The data arrangement is shown in the light green boxes in Figure \ref{fig:netarc}. Before feeding the data into the first \ac{LSTM} layer, the time-dimension is pulled into the batch dimension, which means that all time-steps are processed independently by the first layer, while the second layer processes all frequency bins independently. This simple change enables capturing spectral correlations and gives rise to state-of-the-art multi-channel speaker extraction and enhancement performance as shown in \cite{tesch2022tasl}.

\subsection{Directional conditioning}\label{subsec:condition}

The right part of Figure \ref{fig:netarc} shows the proposed conditioning mechanism, which enables flexible steering of the filter, which was not possible before. The input is a one hot encoding of the target steering direction. The yellow box shows the dimension for a two degree angle resolution, which results in $180$ possible steering directions. Two linear layers are used to map the one-hot encoded input to a dimension that matches in the number of \ac{LSTM} units, which we set to $256$ for the first and $128$ for the second layer. The encoded inputs are then used as initial state for the forward and reverse direction of the bi-directional \ac{LSTM} layers. 

This conditioning mechanism, also used by Vinyals et al. \cite{vinyals2015initstate} for image caption generation, introduces only little overhead as no explicit fusion of input and condition is required. Furthermore, in contrast to \cite{jenrungrot2020coneofsilence}, which is the only other conditioning scheme for steering a DNN-based filter that we are aware of, it does not make a far-field assumption and can thus easily be trained also for larger microphone distances and/or close speakers.

\section{Experimental Setup}\label{sec:expsetup}

\subsection{Datasets}
\begin{figure}[tb]
\begin{tabular}{l l}
\parbox{0.55\linewidth}{
\includegraphics{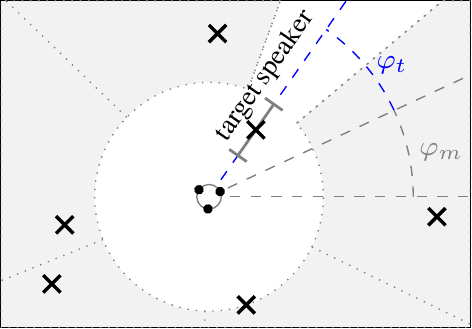}
}
&
\parbox{0.35\linewidth}{
\footnotesize
\begin{tabular}{lr}\toprule
        \multicolumn{2}{c}{\normalsize Room characteristics}\\\midrule
         Width & $2.5-5$ m\\
         Length & $3-9$ m\\
         Height & $2.2-3.5$ m\\
         T60 & $0.2 - 0.5$ s\\\bottomrule
    \end{tabular}
}
\end{tabular}
\vspace{-0.5em}
    \caption{Illustration of the simulation setup. The target source is located on the dashed blue line at a random angle $\varphi_t$ relative to the microphone orientation in the room described by $\varphi_m$. Five interfering sources are placed in the gray area (one per segment). Room properties are uniformly sampled from the given ranges.}
    \label{fig:dataset}
    \vspace{-0.5em}
\end{figure}

We generate a simulated dataset using \texttt{pyroomacoustics} \cite{scheibler2018pyroomacoustics}, which implements the source-image model \cite{allen1979image}. For each sample, we randomly select width, length, height and reverberation time %
from the value ranges given in Figure \ref{fig:dataset}. The left side of Figure \ref{fig:dataset} shows an illustration of the geometric setup of our speaker extraction task. We use a circular microphone array, which has three omni-directional microphones and a $10$~cm diameter. The microphone array is placed at a random location for each example, but with at least one meter distance to the walls and at a fixed height of $1.5$~m above the floor. For each sample, the microphone array is randomly rotated by $\varphi_m\in [0^\circ, 360^\circ]$ as indicated by the dashed gray lines. 

\subsubsection{Fixed target speaker location}\label{subsec:fixedlocation}
The unconditioned joint non-linear filter in \cite{tesch2022tasl, tesch2022interspeech} learns to steer towards a specific direction based on a fixed target location in the dataset. That is, the target speaker is located in the same direction relative to the microphone orientation (position and rotation) in all samples. In Figure \ref{fig:dataset}, for example, the target speaker is located at a $\varphi_t = 30^\circ$ angle as indicated by the dashed blue lines. The distance between the microphone array and the target speaker ranges from $30$ cm to $1$ m. The height of the speakers are sampled from a normal distribution with mean $1.6$~m and standard deviation $0.08$ m. Five interfering sources are placed in the gray area, each of them at least $1$ m away from the microphone array and one per segment as illustrated by the dotted gray lines. As indicated by the white area, we leave a side-room of $15^\circ$ on either side of the target speaker free of interfering sources. %

For training the joint non-linear filter, we generate $6000$ training examples at $16$ kHz sampling frequency with the target speaker located at the chosen direction $\varphi_t$. The clean speech utterances are selected from the WSJ0 dataset respecting its train, test and validation split. The SNR (target speech vs mixture of interfering speakers) of the generated samples distributes in the range from $-14$~dB to $0$~dB. For validation and testing, we generate $1000$ and $600$ utterances respectively.

\subsubsection{Variable target speaker location}\label{subsec:variablelocation}

To train a joint non-linear filter that can flexibly steer towards a selected direction, we create a dataset with a variable target speaker location. For this, we discretize the target speaker location $\varphi_t\in [0^\circ, 360^\circ]$ using a $2^\circ$ resolution, which results in $180$ target speaker locations in the training dataset. We generate a dataset with $300$ utterances per direction, which results in a total of $54000$ training examples. The validation set has $15$ examples per direction. %

\subsubsection{Multiple target speakers (speech separation)}\label{subsec:separation}

In addition to the speaker extraction task, we also evaluate on a speech separation task with multiple target speakers to investigate the spatial selectivity of the filter. The speakers are placed at a distance of $0.8$ to $1.2$ m away from the microphone array. For speaker angle sampling, we split the circle in as many segments as there are speakers and uniformly place each speaker in one of the segments. Consequently, the speaker angles are likely to not lie on the $2^\circ$ grid used in training. A minimum angluar distance of $10^\circ$ is enforced for sources in neighboring segments. We use $1800$ utterances with two, three and five mixed speakers for evaluation.

\subsection{Training details}
The joint non-linear filters are trained based on an $\ell_1$ loss \cite{tolooshams2020cadunet}, i.e.,
\begin{equation}\label{eq:loss}
    L(s, \hat{s}) = \alpha\norm{s-\hat{s}}_1 + \norm{|S|-|\hat{S}|}_1, 
\end{equation}
with $\alpha$ set to $10$ to approximately equalize the contribution of time and frequency-domain loss terms. We train using the Adam \cite{KingmaB2015Adam} optimizer with an initial learning rate of 0.001 and reducing the learning rate by a factor of $\gamma=0.75$ every 50 epochs. We train %
with a maximum of 300 epochs using a batch size of eight and select the best network based on the validation loss.
For computing the \ac{STFT}, we use $32$ ms windows with $50\%$ overlap and a $\sqrt{\text{Hann}}$ window for synthesis and analysis.

\section{Results: speaker extraction} \label{sec:results}

\subsection{Fixed geometry vs conditional training}
\begin{table}[]
    \centering
    \begin{tabular}{lcccccc}\toprule
    & $0^\circ$& $15^\circ$ & $30^\circ$ & $60^\circ$ & $90^\circ$ & $120^\circ$\\\midrule
        JNF (fixed) &  1.38 & 1.36 &1.34 & 1.37 &1.38 & 1.39\\
        EaBNet\cite{li2022eabnet} (fixed) & 1.16 & 1.15 &1.19&1.20&1.18&1.19\\\arrayrulecolor{black!30}\midrule
        JNF (proposed)  & 1.38 & 1.36 & 1.35 & 1.36 & 1.37 & 1.39\\
        JNF (CoS \cite{jenrungrot2020coneofsilence}) & 1.26 & 1.27 & 1.25 & 1.20 & 1.26 & 1.25\\\arrayrulecolor{black}\bottomrule
    \end{tabular}
    \caption{$\Delta$POLQA scores for a fixed training scheme (re-training filter for each angle ($\varphi_t$) with 6000 examples per direction) in the upper part and the filter conditioned on the given direction in the bottom part. Thus, all results in the bottom rows are obtained with the same non-linear filter, which has been trained with 300 examples per direction.}
    \vspace{-0.5em}
    \label{tab:polqaresults}
\end{table}
\begin{figure*}[htb]
    \centering
    \includegraphics{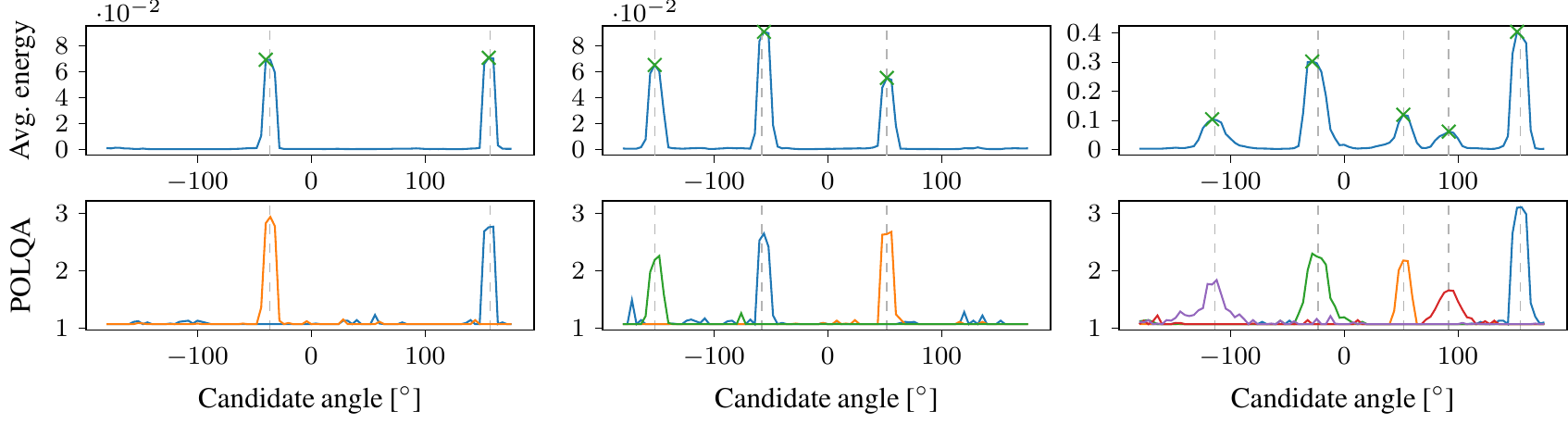}
    \vspace{-1em}
    \caption{Examples for blind speaker separation and localization for a mixture of two, three and five speakers using non-linear filters steered in all candidate directions. The vertical dashed gray lines indicate the true positions of the speakers and the green cross marks the estimates speaker location based on the energy peaks in the results.}
    \label{fig:peaks}
    \vspace{-0.5em}
\end{figure*}
Our first experiment compares the speech extraction performance of a filter trained for a fixed speaker location and a filter that has been trained for variable target speaker locations using the proposed directional conditioning method (Section~\ref{subsec:condition}). The first row of Table~\ref{tab:polqaresults} shows the \ac{POLQA} \cite{polqa2018} \ac{MOS} improvement for six joint non-linear filters, each trained on its own dataset with the target speaker placed at the same respective angle in all 6000 training samples. The improvement performance is very similar for each tested angle, which means that the filters can learn to steer in every direction equally well. As can be seen by the comparison with the Embedding-and-Beamforming Network (EaBNet) in the second row, the learned filters deliver very good state-of-the-art performance. A detailed comparison of more architectures for the $0^\circ$ fixed case can be found in \cite{tesch2022tasl}. 

In contrast, all results displayed in the third row of Table \ref{tab:polqaresults} have been obtained with the same non-linear filter trained on the dataset with variable speaker locations, and conditioned on the respective target angle using the approach proposed in Section \ref{subsec:condition} to obtain the result. As before, we do not observe any major deviations for the different angles and, more importantly, we also do not see a performance degradation in comparison with the non-linear filters in the first row that have been explicitly trained to focus on a fixed spatial location. This is quite remarkable considering that it is a network with only three layers, which is now capable of learning not only one spatial filter but 180 with much fewer (300 instead of 6000) training examples per direction. We also compare with the conditioning mechanism proposed in the cone-of-silence (CoS) paper \cite{jenrungrot2020coneofsilence}. For a fair comparison, we use the same network for the non-linear filter and only replace the conditioning mechanism \cite{jenrungrot2020coneofsilence} as follows: using knowledge of the array geometry, the channels of the input signal are shifted such that the signals arriving from a given target direction should align according to a far-field assumption. To allow for fractional time-shifts, we perform the alignment in the frequency domain. The idea is that the network learns to extract the speaker signal, which is phase-aligned in the input. Given the results in Table \ref{tab:polqaresults}, we find that this seems to be a valid cue for extracting the right target, but that it performs approximately 0.1 POLQA score worse than our proposed direct conditioning. We assume that this is mainly related to the limiting far-field assumption in \cite{jenrungrot2020coneofsilence}. 

\subsection{Spatial selectivity of the steered filter}
Next, we examine the spatial sectivity of the steered filter. Figure~3 shows examples for mixtures of two, three or five speakers. We evaluate the filter on the noisy mixture, generated as described in Section~\ref{subsec:separation}, conditioned on a set of candidate locations using a $4^\circ$ resolution. For each candidate location, we obtain an estimate of the signal arriving from that particular direction. The plots in the top row show the average segmental energy for this resulting signal $\hat{s}$. We compute the average energy for non-overlapping segments of $10$ ms length, in which speech is active, defined analogous to the segmental SNR in \cite{gerkmann2012noisepower}. 

The vertical dashed gray lines indicate the true locations of the speakers in the mixture. In particular for mixtures of two and three speakers, we observe distinct peaks of the energy at the target speaker locations. The small width of the peaks shows high spatial selectivity of the learned filter and proves that it can be steered very accurately towards a specific location. For a mixture of five speakers, we can still see peaks that correspond to the speaker locations, but with a greater width. Likely this is due to the much increased difficulty by a larger overlap of the signals in the time-frequency plane and more reflections arriving from all directions. Still, the POLQA results in the bottom row of Figure \ref{fig:peaks} show that even five speakers can be separated quite well by the spatially selective filters, which is remarkable given the difficulty of the problem. Audio examples can be found on our website \footnote[1]{\footnotesize\url{https://uhh.de/inf-sp-spatially-selective}}. 

\FloatBarrier
\newcolumntype{m}[1]{D{:}{\pm}{#1}}
\newcolumntype{C}{ @{}>{${}}c<{{}$}@{} }
\begin{table}[htb]
    \centering
\begin{tabular}{l *3{rCl} }
    \toprule
    \multirow{2}{*}[-2pt]{\# speakers} & \multicolumn{9}{c}{mean angular error [$^\circ$]}\\\cmidrule{2-10}
    & \multicolumn{3}{c}{proposed} & \multicolumn{3}{c}{CoS\cite{jenrungrot2020coneofsilence}} & \multicolumn{3}{c}{SRP-PHAT\cite{dibase2000srpphat}}\\\midrule
    2 & 2.17   & \pm& 0.13                & 3.72  & \pm & 0.46    & 17.74 & \pm & 1.04\\
    3 & 2.47  & \pm & 0.16                & 3.72 & \pm & 0.36            & 20.47 & \pm & 0.79\\
    5 & 3.50 & \pm & 0.21               & 5.85   & \pm & 0.35      & 25.62 & \pm & 0.59\\ \bottomrule
\end{tabular}
    \caption{The speaker localization accuracy for mixtures of two, three and five speakers in a reverberant room. We report the mean angular error and the 95\% confidence interval.}
    \label{tab:loc_acc}
\end{table}
The green crosses in the top row mark the estimated speaker locations that have been found using \texttt{scipy.signal.find\_peaks}. We normalize the highest peak to 1 and initially use a prominence of 0.009 and a height of 0.05, which are decreased until enough peaks have been identified. We then merge close-by peaks less then $12^\circ$ apart and with similar height, which are likely to correspond to the same speaker. If more peaks than speakers are detected, we select the highest peaks. Comparing the distance of the green crosses to the dashed gray line with respect to the x-dimension, shows that the location of the target speakers can be estimated from the steered filter's results quite accurately. In Table \ref{tab:loc_acc}, we compare the estimated speaker locations using a $4^\circ$ resolution with the true speaker locations on $1800$ mixtures. For two speakers, the average error is only $2.2^\circ$ including an average quantization error of $1^\circ$ (as the speaker locations are not limited to the $4^\circ$ test grid). The error increases for more speakers mainly due to a higher number of errors in the peak-finding heuristic and is still fairly accurate considering the difficulty of the task using only three microphones in a reverberant room. This difficulty is also visible from the fact that the classic SRP-PHAT algorithm \cite{dibase2000srpphat} is not able to solve the problem in most cases even for only two speakers. As for the extraction task in Table~\ref{tab:polqaresults}, we observe that our proposed conditioning scheme outperforms the baseline CoS approach in all configurations.

\section{Conclusion}
In this paper, we have presented a simple but very effective conditioning mechanism to train a non-linear filter that can be steered in any direction of choice. The conditioning is performed by modifying the initial state of the \ac{LSTM} layers in the non-linear filter and, thus, introduces only minimal overhead, while achieving the same state-of-the-art performance as a filter with fixed look-direction and also outperforming the baseline cone-of-silence approach. We show that the resulting spatially selective filters can be used for speech separation with an arbitrary number of speakers and can also be employed for accurate multi-speaker localization.

\bibliographystyle{IEEEtran}
\bibliography{bib}

\end{document}